# Sustainable Software Development for Next-Gen Sequencing (NGS) Bioinformatics on Emerging Platforms

Shel Swenson[1,2], Yogesh Simmhan[1], Viktor Prasanna[1], Manish Parashar[3],
Jason Riedy[2], David Bader[2] and Richard Vuduc[2]

[1] University of Southern California, [2] Georgia Institute of Technology, [3] Rutgers University

## Background and Motivation

The advent of Next Generation Sequencing (NGS) technology more than half a decade ago produced a deluge of DNA sequence data by completely altering the scale at which sequence data is generated and dramatically reducing sequencing costs. Both the increase in sheer volume of data being produced —fed by the continually increasing throughput and decreasing costs of sequencing technologies— and the unique and evolving characteristics of NGS data itself continue to fuel the development of countless new analysis tools.

At the same time, the pace of development of novel hardware platforms has seen a resurgence. As the 'Moore's Law' of computational progress plateaus in single-chip performance, we have seen the emergence (or renewed focus) on accelerated platforms such as General Purpose Graphics Processing Units (GPGPUs), Field Programmable Gate Arrays (FPGAs), massively multi-threaded and many-core processors. These accelerators supplement the democratized access to computing infrastructure made possible through Cloud and commodity clusters. These accelerators offer several benefits over traditional single processor machines, including a large memory bandwidth, massive parallelism, elastic scaling, and a small energy footprint, although each technology presents its own trade-off between these benefits. Accelerators also becoming more affordable, to the extent that GPGPUs are routinely found in scientific workstations and are being embedded within processors.

The rapid pace of hardware innovation, however, has not been adequately supported by accessible middleware and software kernels that make these cyber-infrastructure more widely usable. The challenges stretch from the learning curve of specialized programming models and abstractions to the need for novel algorithms that leverage these accelerators effectively. Consequently, the impact of accelerators on scientific and 'Big Data' computing has not been as transformative as the potential of them. Indeed, for a rapidly changing domain such as NGS where oftentimes data lies unused due to the lack of scalable analytical capabilities, the use of accelerated cyber-infrastructure supported by sustainable software infrastructure is vital to spur scientific innovation.

Unlocking the inner-workings of genomic sequences is at the heart of many grand challenge problems in biological research. The growth available sequence data engendered by NGS technology provides promise of solutions to many of these problems; however, the development of sequence analysis techniques has not kept pace. Bridging this gap holds possibilities for disruptive progress toward multiple grand challenges and offers unprecedented opportunities for software innovation and research. Moreover, the complexity of the software pipelines involved in most NGS-enabled analyses creates an urgent need for *sustained community software stacks* that can enable scientific innovation.

We argue that NGS-enabled applications need a critical mass of sustainable software to benefit from *future platforms'* transformative potential.  These Future Computing Platforms(FCP) include GPGPUs, FPGAs multi -core and-threaded processors, and Cloud and commodity clusters. Accumulating the necessary critical mass will require leaders in computational biology, bioinformatics, computer science, and computer engineering work together to identify

1. *core opportunity areas* where FCP can make a significant impact in solving one or more grand challenges in the domain,
2. *critical software infrastructure*, both existing and missing, in the form of tools, models, algorithms, libraries, frameworks and middleware necessary to map application needs to the platform capabilities, and
3. *software sustainability challenges* which must be addressed to sustain innovation and accelerate scientific progress.

Furthermore, due to the quickly changing nature of both bioinformatics software and accelerator technology, we conclude that creating *sustainable accelerated bioinformatics software* means constructing a *sustainable bridge* between the two fields. In particular, sustained collaboration between domain developers and technology experts is needed to develop the accelerated kernels, libraries, frameworks and middleware that could provide the needed flexible link from NGS bioinformatics applications to emerging platforms.

## Understanding the unique challenges in NGS bioinformatics

To begin identifying these core opportunity areas, critical software infrastructure, and software sustainability challenges, we organized a one-day workshop on "Future computing platforms to accelerate Next-Gen Sequencing (NGS) applications" held May 19th 2013, in Boston. This workshop was part a broader multi-institutional project [1] (supported by the NSF BIO and CISE directorates) focused on supporting sustainable software for data-intensive and inter-disciplinary problems that may be enabled by graph algorithms. Graph algorithms are at the heart of many important bioinformatics, epidemiology, and biomedical applications, including those enabled by NGS data, and map well to FCP, making NGS-enabled application areas a natural fit for our project's overall goals. The workshop included more than a dozen interdisciplinary scientist from biology and computer science (CS) including representatives from iPlant[2], Galaxy[3], and Big Data for NGS applications[4], as well as designers and developers of several widely used NGS-enabled software applications. The main objectives were:

- Identify grand challenge problems in NGS bioinformatics with the best opportunities for emerging platforms, and specific algorithms of broad use and high-impact for NGS to accelerate on these platforms.

- Identify support mechanisms needed by domain researchers to accelerate scientific progress.

In line with these goals, participants produced potential usecases, identified other applications or groups could benefit from these efforts, and suggested support mechanisms to help domain researchers utilize accelerator technologies [5]. Here we discuss a key insight gained from the workshop.

All participants agreed that for accelerated algorithms to be broadly used in the BIO community, tighter integration between domain scientists and technology experts will be key. The factors that create this necessity for increased integration are that in this domain,
- problems change rapidly,
- analyses typically involve complex of the software pipelines,
- there is an incredibly diverse community of users, and
- software users and developers rely on trusted (and user-friendly) packages and development platforms.

The overwhelming majority of software available, including those most commonly used, are serial and do not take advantage of accelerator platforms. There are some efforts to use these platforms but these approaches may not scale to the demand, and are rarely portable to other applications or even to other programs designed for the same application. Therefore, these accelerated algorithms cannot be used outside of the specific software pipeline for which they were developed. Taken together, these factors lead us to conclude that expert developers on emerging platforms must work closely with designers and users of trusted software development platforms to fashion the libraries and frameworks that can work seamlessly with those trusted packages and environments.

## Strategies for sustainable development of NGS-enabled bioinformatics software on emerging platforms

The workshop and subsequent investigations elucidated promising avenues in each of our three lines of inquiry (core opportunity areas, critical software infrastructure, and software sustainability challenges), each of which is critical in developing sustainable software on emerging platforms in this domain. We will discuss here the findings with respect to critical software infrastructure, and software sustainability challenges. (A discussion of the core opportunity areas can be found in the workshop report [5].)

**Software infrastructure.** In addition to the need for accessible middleware and accelerated software kernels for core graph algorithms, we must also address hurdles in the full software stack: from access to specific FCPs, to accelerated graph algorithm libraries, to development platform and to distribution channels, to end user. The usability of accelerated software kernels depends on the entire software stack, and thus must be developed taking all these layers into account to achieve the portability necessary to achieve a broad user base.

For example, assembling transcriptomes directly from sequence reads without the use of a reference genome (de novo assembly) is of critical importance for any group studying gene expression in non-model organisms, and was identified as a core opportunity area during the workshop. Of the popular de novo transcriptome assembly software packages, "the majority of participants agreed that Trinity [6] was the currently the best target for development of FCP implementations, but given the rate at which assemblers come in and out of fashion, development of *extensible* implementations of the core graph algorithms is key."[5] Participants further identified efforts such as the Galaxy Project and iPlant Collaborative as potentially instrumental in connecting domain users with accelerated applications, and recommend pursuing these potential synergies. Subsequent discussions with bioinformatics tools developers, suggest supercomputing centers (eg.,Texas Advanced Computing Center or Pittsburg Supercomputing Center) or broader efforts such as XSEDE [7] could make up the FCP layer of the software stack, and should also be pursued.

**Supporting software sustainability.** Enabling wide-scale use of accelerated graph algorithms by NGS bioinformatics software developers, in such a way that the software and incorporation of FCP will become self-sustaining, requires developing a wide base of users. In addition to choosing the right algorithms and potential software stacks to support, our workshop participants felt that outreach and training to demonstrate the advantages of using accelerated algorithms will be key. "Further, that the primary target for these interactions should be software developers who are tightly connected with the bio community. By creating and supporting a strong user (developer) community the software and incorporation of FCP will become self-sustaining. Suggestions as to how to maintain this developer community were to bring them together on a regular basis via, for example, yearly training workshops and hackathons." [5]

## References

1. Accelerating Grand Challenge Data-intensive Problems using Future Computing Platforms. NSF Software Institute Conceptualization, Award #1216898, #1216504, and #1216696. *http://future-compute.usc.edu.*
2. iPlant Collaborative. *http://www.iplantcollaborative.org.*
3. Galaxy Project *http://galaxyproject.org.*
4. Genomes Galore - Core Techniques, Libraries, and Domain Specific Languages for High-Throughput DNA Sequencing. NSF BIGDATA, Award #1247716.
5. Future computing platforms to accelerate Next-Gen Sequencing (NGS) applications workshop report. http://future-compute.usc.edu/index.php/NGS_Workshop.
6. M.G. Grabherr, et. al. Full-length transcriptome assembly from RNA-seq data without a reference genome. *Nature Biotechnology*. 29(7):644-52, 2011.
7. XSEDE. *https://www.xsede.org.*